\documentclass[12pt,a4wide]{article}
\usepackage{amsmath,amssymb,amsfonts}

\newcommand{\eqa}{\begin{eqnarray}}
\newcommand{\ena}{\end{eqnarray}}
\begin{document}
\begin{center}
{\large {\bf Interpretations of Cosmological Spectral Shifts}}
\end{center}
\begin{center}
Dag {\O}stvang \\
{\em Department of Physics, Norwegian University of Science and Technology
(NTNU) \\
N-7491 Trondheim, Norway}
\end{center}
\begin{abstract}
It is shown that for Robertson-Walker models with flat or closed space 
sections, {\em all} of the cosmological spectral shift can be attributed to the
non-flat connection (and thus indirectly to space-time curvature). For 
Robertson-Walker models with hyperbolic space sections, it is shown that 
cosmological spectral shifts uniquely split up into ``kinematic'' and 
``gravitational'' parts provided that distances are small. For large distances 
no such unique split-up exists in general. A number of common, but incorrect 
assertions found in the literature regarding interpretations of cosmological 
spectral shifts, is pointed out. \\
\end{abstract}
\topmargin 0pt
\oddsidemargin 5mm
\renewcommand{\thefootnote}{\fnsymbol{footnote}}
\section{Introduction}
Although there is in general no dispute about actual predictions coming from
universe models based on General Relativity (GR), {\em interpretations} of the 
nature of the cosmic expansion/contraction and cosmic spectral shifts predicted
by these models, have on the other hand been subject to some lengthy 
controversy. (See e.g., [1] and references therein.) The controversial question
is, when pulses of electromagnetic radiation are emitted and received between 
``fundamental observers'' (FOs) following the cosmic fluid (with no peculiar 
motions); what is the nature of the resulting spectral shifts?

What is completely {\em uncontroversial} is the fact that the ratio of the 
observed and emitted wavelengths ${\lambda}_{\rm obs}$ and ${\lambda}_{\rm em}$, 
respectively, is related to the ratio of the cosmic scale factors at 
observation and emission $a_{\rm obs}$ and $a_{\rm em}$, respectively, and the 
cosmological spectral shift $z$ via the formula 
${\lambda}_{\rm obs}/{\lambda}_{\rm em}=a_{\rm obs}/a_{\rm em}=1+z$. However, the
controversial part of said question is to what extent, if any, such 
cosmological spectral shifts can be interpreted as Doppler shifts in flat 
space-time.

One school of thought claims that, since the Equivalence Principle (EP) says 
that space-time is locally flat, the nature of cosmological spectral shifts 
must be interpreted as Doppler shifts in flat space-time in the limit where
distances between FOs go to zero. A different view is that, since the FOs are 
at rest with respect to the cosmic fluid (defining an ``expanding frame'') and 
since cosmological spectral shifts are given from the formula shown above 
rather than from the special-relativistic Doppler formula, cosmological 
spectral shifts should in principle have nothing to do with Doppler shifts, 
even for arbitrarily small distances. These interpretations are known as the 
``kinematic'' (in the narrow sense of the word) and the ``expanding space'' 
interpretations, respectively. Note that for curved space-times, there is no 
general agreement between proponents of the ``expanding space'' interpretation 
on whether or not cosmological spectral shifts should be entirely attributed to 
space-time curvature. However, some authors claim this and thus that 
cosmological spectral shifts should be interpreted as some sort of 
``gravitational'' spectral shifts whenever space-time is not flat.

At first glance, at least for small distances, the difference between these 
interpretations seems to be the rather trivial matter of describing the same 
physics using different frames. Thus, to some people it would seem reasonable 
to proclaim both interpretations valid for small distances and just 
representing equivalent points of view. However, before such a solution is 
endorsed, it must be established that the various interpretations are 
mathematically consistent. But it turns out that they aren't, since 
interpretations may be associated with geometrical restrictions. In particular, 
in this paper we show that the ``kinematic'' interpretation is in general 
mathematically inconsistent with the geometry of the Robertson-Walker (RW) 
models, so that this interpretation is not valid generally. On the other hand, 
we show that what is crucial for interpreting spectral shifts in the RW-models 
is not the mere existence of an ``expanding frame'', but rather how this frame 
relates to space-time curvature. And as we shall see in the next section, this 
relationship differs between types of RW-models. That is why the
``expanding space'' interpretation is not useful in general, either.

Besides, since proponents of the ``expanding space'' interpretation do not in 
general agree on how it is related to space-time curvature, even for small 
distances, it will be more clarifying to talk about spectral shifts due to 
space-time curvature rather than the ``expanding space'' interpretation.  
However, there is a common point of view claiming that spectral shifts cannot 
really be an effect of space-time curvature since, unlike the equation of 
geodesic deviation, the geodesic equation does not contain components  of the
Riemann tensor, but only connection coefficients. It is true that, unlike tidal
effects, spectral shifts cannot be a {\em direct} effect of space-time 
curvature, i.e., representing operational measures of it. But it is indeed 
possible that spectral shifts may represent an {\em indirect} effect of 
space-time curvature. 

To see this, define (e.g., via coordinate-parametrisation) specific world lines
in a (sufficiently small) region of some (curved) space-time geometry. 
Calculate spectral shifts obtained by photon signalling between observers 
moving along the chosen world lines. Now replace the metric with its 
flat space-time form in the region, holding the chosen world lines and the 
coordinate system fixed. Calculate spectral shifts again, but now with the flat
space-time geometry. If the results are different for the two cases, this must 
certainly be due to space-time curvature via the non-flat connection. 

In particular, it may be possible that the latter calculation will yield no 
spectral shift at all. In such a case, it is rather obvious that the spectral 
shift should unambiguously be interpreted as purely ``gravitational'', i.e., 
as an effect purely due to space-time curvature. We show in the next section 
that this situation arises for any RW-model with flat or spherical space 
sections with the FOs playing the role as ``preferred'' observers 
operationally defining cosmological spectral shifts.
\section{``Kinematic'' and ``gravitational'' spectral shifts}
To understand what is actually meant by ``kinematic'' and ``gravitational'' 
spectral shifts in context of the RW-models, it is necessary to define these 
concepts mathematically. Such definitions should be formulated together with
a recipe for spectral shift split-up into ``kinematic'' and ``gravitational'' 
parts. It would perhaps seem natural to insist that said definitions must be
based on a general spectral shift split-up coming from some geometrical 
procedure being valid for all RW-models. However, it is shown in this section 
that such an approach cannot be justified if the definition of ``kinematic'' 
spectral shift is required to be based on the definition of spectral shifts in 
Special Relativity (SR). Abandoning this requirement is certainly 
possible, but then the definitions of ``kinematic'' and ``gravitational'' 
spectral shifts will be only formal and misleading, and thus not very useful 
for interpretations.

The mathematical framework considered in this paper is given by the usual 
4-dimensional semi-Riemann manifold $({\cal M},{\bf g})$. In addition it is 
required that (at least a subset of) $({\cal M},{\bf g})$ can be foliated into 
a continuous sequence of 3-dimensional spatial hypersurfaces ${\cal S}(x^0)$ 
parameterized by a time function $x^0$. The {\em fundamental observers} are 
``preferred'' observers defined from the foliation by the criterion that their 
world lines are everywhere continuous and orthogonal to ${\cal S}(x^0)$. The 
choice of foliation (and thus of time coordinate) is required to be 
unambiguously made from purely geometrical selection criteria.

Since this paper is about the RW-manifolds, the analysis presented here is 
restricted to one specific choice of selection criteria. This specific choice 
of selection criteria picks out space-time manifolds that can be foliated into 
a set of hypersurfaces such that the spatial geometry is everywhere isotropic 
and homogeneous. Moreover, the unit normal vector field to the hypersurfaces 
should not be a (time-like) Killing vector field. This last criterion excludes
static manifolds with topology ${\cal S}{\times}{\bf R}$ equipped with 
a foliation determined from the product topology (here $\cal S$ is one 
of the space geometries ${\bf R}^3$, ${\bf S}^3$ or ${\bf H}^3$). These 
selection criteria uniquely yield the RW-manifolds each equipped with the 
prescribed ``preferred'' foliation, determining ``preferred'' hypersurfaces. 
(In the next section, we will also consider other foliations of ``open'' 
RW-manifolds than the ``preferred'' ones as useful for specific calculations.)

Given $({\cal M},{\bf g})$ and a foliation of it into spatial hypersurfaces
${\cal S}(x^0)$, spectral shifts obtained by exchanging light signals between 
nearby FOs are unambiguously determined from the space-time geometry. Moreover,
this holds irrespectively of any particular choice of field equations, so it 
is not necessary to assume the validity of the GR field equations. 
Consequently, the results obtained in this paper depend only on the geometry 
of space-time with no extra assumptions. In particular, no particular 
relationship between geometrical quantities and matter sources is assumed to 
hold. Nevertheless, we will by convention call spectral shifts entirely due to 
space-time curvature for ``gravitational''spectral shifts. (See Definition 1
below for such a situation.)

To calculate spectral shifts in general, there exists a simple geometric 
procedure, as first pointed out by Synge. That is, imagine a pulse of 
electromagnetic radiation being emitted at some given event and subsequently 
observed at some other given event. Then, by parallel-transporting the 
4-velocity ${\bf u}_{\rm e}$ of the emitter along the null curve connecting the
given events, the parallel-transported 4-velocity of the emitter can be 
projected into the local rest frame of the observer. This yields a 3-velocity 
that can be inserted into the special-relativistic Doppler formula to give the 
desired spectral shift. (For the full mathematical details of this procedure, 
see [2].) This procedure works for any relativistic space-time (and even for 
cases where the space-time geometry is not semi-Riemann [3]), and implies 
that it is always possible to interpret spectral shifts as due to the Doppler 
effect in {\em curved} space-time, without any geometrical restrictions 
whatsoever. Moreover, this procedure illustrates that what is relevant for 
calculating spectral shifts are the connection coefficients, since these enter 
into the mathematical expression for parallel-transport. Any non-zero values 
of the connection coefficients may arise due to the choice of coordinates, a 
non-flat connection, or both.

For any RW-manifold, given some coordinate system covering (part of) it, the 
connection coefficients relevant for spectral shifts obtained from photon 
signalling between FOs are uniquely determined from the evolution with time 
of the spatial geometry ${\bf h}$ of the ``preferred'' foliation in the 
direction of the unit vector field ${\bf n}$ normal to the hypersurfaces. That 
is, the relevant quantity is given from the {\em extrinsic curvature tensor} 
${\bf K}$ defined by (in component notation using a general coordinate system 
${\{}x^{\mu}{\}}$ and using Einstein's summation convention, see, e.g., ref. 
[4], p. 256)
\eqa
K_{{\mu}{\nu}}{\equiv}{\frac{1}{2}}{\pounds}_{\bf n}h_{{\mu}{\nu}}
={\frac{1}{2}}{\Big (}h_{{\mu}{\nu}},_{\alpha}n^{\alpha}+
h_{{\alpha}{\nu}}n^{\alpha},_{\mu}+h_{{\mu}{\alpha}}n^{\alpha},_{\nu}{\Big )},
\ena
where ${\pounds}_{\bf n}$ denotes the Lie derivative in the 
${\bf n}$-direction and where a comma denotes a partial derivative.
Using a spherically symmetric hypersurface-orthogonal coordinate system 
${\{}x^0,{\chi},{\theta},{\phi}{\}}$ where $n^{\mu}=(1,0,0,0)$, the 
metrics of the RW-manifolds (equipped with the ``preferred'' foliation) take 
the form
\eqa
{ds}^2=-(dx^0)^2+a^2(x^0){\Big (}d{\chi}^2+
{\Sigma}^2({\chi})d{\Omega^2}{\Big )}, \qquad 
d{\Omega}^2{\equiv}d{\theta}^2+{\sin}^2{\theta}d{\phi}^2,
\ena
where $a(x^0)$ is the scale factor of the hypersurfaces and where
\eqa
{\Sigma}({\chi})=
\left\{
\begin{array}{ll}
{\sin}{\chi}
& \rm{for{\ }hypersurfaces{\ }with{\ }spherical{\ }geometry,} \\ [1.5ex]
{\chi}
& \rm{for{\ }hypersurfaces{\ }with{\ }flat{\ }geometry,} \\ [1.5ex]
{\sinh}{\chi}
& \rm{for{\ }hypersurfaces{\ }with{\ }hyperbolical{\ }geometry.}
\end{array}
\right.
\ena
Using the form (2) of the metric, equation (1) for the extrinsic curvature
of the ``preferred'' hypersurfaces (embedded into the RW-manifolds) takes the 
form (with ${\dot a}{\equiv}{\frac{da}{dx^0}}$)
\eqa
K_{{\mu}{\nu}}={\frac{1}{2}}{\frac{\partial}{{\partial}x^0}}h_{{\mu}{\nu}}
={\frac{1}{c}}H(x^0)h_{{\mu}{\nu}}, \quad H(x^0){\equiv}
{\frac{{\dot a}}{a}}c,
\ena
where $H(x^0)$ is the Hubble parameter. In a hypersurface-orthogonal coordinate
system (such as used in equation (2)), the nonzero components of the spatial 
metric ${\bf h}$ can be found directly from the spatial part of the space-time 
metric ${\bf g}$, yielding the nonzero components of ${\bf K}$ from equation 
(4). Note that ${\bf K}$ is a tensor field on space (since the scalar product
${\bf K}{\cdot}{\bf n}={\bf 0}$), and that 
$H(x^0)={\frac{c}{3}}K^{\mu}_{{\ }{\mu}}$ is a scalar field (constructed from the
``preferred'' foliation). This means that $H(x^0)$ is not a 
coordinate-dependent quantity, despite the fact that the relationship between 
$H(x^0)$ and the connection coefficients certainly is.

Now it turns out that there exists an expression for the intrinsic Riemann
curvature tensor ${\bf P}$ of the hypersurfaces in terms of the space-time 
Riemann curvature tensor ${\bf R}$ and the extrinsic curvature tensor 
${\bf K}$. This is the well-known Gauss equation (see, e.g., ref. [4], p. 258), 
and in component notation it reads
\eqa
P^{\alpha}_{{\ }{\beta}{\gamma}{\delta}}=
R^{\lambda}_{{\ }{\rho}{\mu}{\nu}}h^{\alpha}_{{\ }{\lambda}}
h^{\rho}_{{\ }{\beta}}h^{\mu}_{{\ }{\gamma}}h^{\nu}_{{\ }{\delta}}
+K^{\alpha}_{{\ }{\delta}}K_{{\beta}{\gamma}}
-K^{\alpha}_{{\ }{\gamma}}K_{{\beta}{\delta}}.
\ena
Moreover, contracting equation (5) twice and using equation (4), we get
\eqa
P=R+2R_{{\alpha}{\beta}}n^{\alpha}n^{\beta}-6H^2/c^2, \quad \Rightarrow \quad
H^2={\frac{c^2}{6}}{\Big [}2G_{{\alpha}{\beta}}n^{\alpha}n^{\beta}-P{\Big ]},
\ena
where $R$ and $P$ are the scalar curvatures of space-time and space, 
respectively, and where $R_{{\alpha}{\beta}}$ and $G_{{\alpha}{\beta}}$ are the 
components of the Ricci tensor and the Einstein tensor on space-time, 
respectively. Equation (6) represents a well-known constraint equation as
part of the initial-value problem applied to the RW-manifolds equipped with the 
``preferred'' foliation.

Now we see from equation (4) that there can be no spectral shift (detected by
photon signalling between FOs) if ${\bf K}={\bf 0}$. Therefore, to make sense
of any ``kinematic'' part of the spectral shift having a similarity with 
spectral shifts in SR, it must be possible to have a limit where the relevant 
part of ${\bf R}$ may be neglected but such that ${\bf K}{\neq}{\bf 0}$. If 
such a limit does not exist, the spectral shift must be entirely due to 
space-time curvature (i.e., ``gravitational''). Whether or not such a limit 
exists can be found from equation (6). (The Weyl tensor vanishes identically for
the RW-manifolds, so the Ricci tensor (or the Einstein tensor) captures all
aspects of space-time curvature.) We thus have the definition
\newtheorem{definition}{Definition}
\begin{definition}
Assume as given a semi-Riemann manifold $({\cal M},{\bf g})$ of RW-type and 
a foliation of it into ``preferred''isotropic and homogeneous spatial 
hypersurfaces ${\cal S}(x^0)$ (with unit normal vector field ${\bf n}$) defined 
from equation (2). Also denote any hypersurface metric by ${\bf h}$ with 
extrinsic curvature tensor ${\bf K}$, intrinsic Riemann curvature tensor 
${\bf P}$ and intrinsic curvature scalar $P$. The space-time Einstein curvature 
tensor is denoted by ${\bf G}$. Then, if it is not possible to have 
$G_{{\alpha}{\beta}}n^{\alpha}n^{\beta}$ arbitrary small independent of $P$ with
${\bf K}{\neq}{\bf 0}$, any spectral shift obtained by photon signalling 
between FOs is entirely due to space-time curvature. 
\end{definition}
If a situation like that described in Definition 1 occurs, any definitions and
spectral split-ups that allow for a non-zero ``kinematic'' spectral shift do 
not make sense, since the correspondence with spectral shifts in SR will be 
lost. This is why the approach of starting with general definitions of 
``kinematic'' and ``gravitational'' spectral shifts valid for any RW-models 
cannot be justified, since, as we shall see, the situation described in 
Definition 1 occurs for {\rm all} RW-models where the ``preferred'' foliation 
consists of flat or spherical hypersurfaces.

To prove that the situation described in Definition 1 occurs for the case of
flat hypersurfaces, it is obvious from equation (6) that it is not possible to 
have a flat RW-manifold with flat spatial sections (i.e.,
${\bf P}{\equiv}{\bf 0}$) and still have $H(x^0){\neq}0$. That is, the 
requirements $P=0$, ${\bf G}={\bf 0}$ mean that equation (6) is satisfied only 
for $H(x^0)=0$. Note that this is not in any way a coordinate-dependent result.
Thus we arrive at the conclusion that to have a RW-manifold with flat space 
sections and at the same time $H(x^0){\neq}0$, space-time must be curved. This 
means that according to Definition 1, {\em spectral shifts observed by 
exchanging photons between FOs in a RW-manifold with flat spatial sections are 
entirely due to space-time curvature.} Since this result holds irrespective of 
distances between FOs, we are forced to interpret the relevant spectral shifts 
as purely ``gravitational'' for all RW-manifolds with flat spatial sections.

A similar result holds for the closed RW-manifolds (with spherical spatial 
sections). In this case $P={\frac{6}{a^2}}>0$, and equation (6) yields that it 
is not possible to have $G_{{\alpha}{\beta}}n^{\alpha}n^{\beta}$ arbitrary small 
independent of $P$ such that $H^2>0$. This result is a consequence of the fact
that it is not possible to foliate Minkowski space-time into hypersurfaces with
${\bf S}^3$-geometry. So, from Definition 1 we have that {\em spectral shifts 
observed by exchanging photons between FOs in a RW-manifold with closed 
(spherical) spatial sections are entirely due to space-time curvature.} We are 
then forced to interpret all relevant spectral shifts as purely 
``gravitational'' for all RW-manifolds with spherical spatial sections as well.

We are thus left with open RW-manifolds foliated into hyperbolical 
hypersurfaces as the only nontrivial case when it comes to interpretations.
In this case $P=-{\frac{6}{a^2}}<0$, so it is indeed possible to choose 
$G_{{\alpha}{\beta}}n^{\alpha}n^{\beta}$ arbitrary small independent of $P$ in equation
(6) together with $H^2>0$ (i.e., ${\bf K}{\neq}{\bf 0}$), so that the situation
described in Definition 1 does not occur. This means that for the case $P<0$, 
it may make sense to define a spectral shift split-up into ``kinematic'' and 
``gravitational'' parts. In particular it is possible to choose 
${\bf G}={\bf 0}$, $P<0$ in equation (6) together with $H^2(x^0)>0$, since 
(part of) Minkowski space-time can be foliated into hypersurfaces with 
${\bf H}^3$-geometry. This special case is the ``empty'' RW space-time (Milne 
model), which is just a subset of Minkowski space-time and thus flat. The line 
element is given by equations (2) and (3) by setting $a(x^0)=x^0$, i.e.,
\eqa
{ds}^2=-(dx^0)^2+(x^0)^2(d{\chi}^2+{\sinh}^2{\chi}d{\Omega^2}).
\ena
In this case it is obvious that the ``kinematic'' interpretation is correct 
since the cosmic expansion is entirely due to the ``preferred'' choice of 
space-time foliation into space and time. That is, by switching to standard 
coordinates $r{\equiv}x^0{\sinh}{\chi}$, ${x^0}'{\equiv}x^0{\cosh}{\chi}$, 
another foliation is chosen and the line element takes the familiar 
Minkowski form expressed in spherical coordinates. This means that, by 
performing a suitable coordinate transformation, it is possible to eliminate 
the connection coefficients altogether. Moreover, the cosmic redshift can be 
found {\em locally} from the speed $w_{\cal F}$ of a FO relative to a local 
observer moving normal to the ${x^0}'=$constant hypersurfaces. We will exploit 
this fact when treating general open models in section 3.

We may now define a ``purely kinematic'' spectral shift as one occuring in a 
RW-manifold foliated into hyperbolical hypersurfaces for situations where the 
difference between a non-flat and a flat connection does not matter for photon 
propagation between nearby FOs. That is, it may be possible that the 
contribution to equation (5) from extrinsic curvature at some epoch $x^0_0$ 
is identical to the contribution to equation (5) from extrinsic curvature of 
a hyperbolic hypersurface with identical geometry but embedded in Minkowski 
space-time. (In such a situation, $H^2$ and $P$ will be identical for the two 
hypersurfaces, meaning that $G_{{\alpha}{\beta}}n^{\alpha}n^{\beta}$ must vanish even 
at the hypersurface embedded in curved space-time in order not to violate 
equation (6).) To find what the latter contribution is, it is 
convenient to use a hypersurface-orthogonal coordinate system as used in 
equation (2). One then finds that the contribution to equation (5) from 
extrinsic curvature depends on ${\dot a}^2(x^0)$ but not on $a(x^0)$. Since 
${\dot a}^2(x^0)=1$ for the Milne model given in equation (7), we have the 
definition:
\begin{definition}
Assume as given a semi-Riemann manifold $({\cal M},{\bf g})$ of RW-type and 
a foliation of it into isotropic and homogeneous spatial hypersurfaces 
${\cal S}(x^0)$ with hyperbolic intrinsic geometry (see equation (2)). Also 
assume the existence of some hypersurface ${\cal S}(x^0_0)$ with spatial metric
${\bf h}(x^0_0)$ and extrinsic curvature tensor ${\bf K}(x^0_0)$ given from 
equation (4) (in a hypersurface-orthogonal coordinate system) with 
${\mid}{\dot a}(x^0_0){\mid}=1$. Then spectral shifts resulting from photon 
signalling between nearby FOs close to ${\cal S}(x^0_0)$ are defined to be 
``purely kinematic'' in the limit where distances between FOs go to zero 
and $x^0{\rightarrow}x^0_0$.
\end{definition}
Definition 2 is based on the fact that for the open RW-models, it may be 
possible to have a situation where the contribution to equation (5) from 
extrinsic curvature at the epoch $x^0_0$ is identical to that of a hyperbolic 
hypersurface with identical geometry in the Milne model at epoch $x^0_0+b$ 
where $b$ is some constant. In such a situation, the relevant connection 
coefficients for the open model at epoch $x^0_0$ will be identical to those 
for the Milne model at epoch ${\tilde x}^0_0$, where the scale factor is given 
by $a({\tilde x}^0)={\tilde x}^0=x^0+b$ and such that 
$a({\tilde x}^0_0)=x^0_0+b=a(x^0_0)$. In the next section, we will give some 
specific examples of open RW-models where this situation occurs.
\section{Spectral shift split-up}
The main result of the previous section was that {\em all} RW-models foliated 
into flat or spherical hypersurfaces satisfy the situation described in 
Definition 1. Therefore, {\em all} the cosmic spectral shift in these models 
must be due to space-time curvature, so that a spectral shift split-up into 
``kinematic'' and ``gravitational'' parts does not make sense for these 
RW-models. On the other hand, we show in this section that for RW-models 
foliated into hyperbolical hypersurfaces, such a spectral shift split-up can be
defined consistently for small distances and in agreement with Definition 2.

To define a split-up of spectral shifts into ``kinematic'' and 
``gravitational'' parts (valid for RW-models with hyperbolical spatial 
sections), it is convenient to change the space-time foliation. Note that the
change of foliation is made because it makes calculations easier and the
correspondence with the Milne model clearer. Note in particular that the FOs
are still being defined as those observers moving normal to the ``preferred''
foliation given from equation (2), and that the defined spectral shift 
split-up applies only to the FOs. Observers moving normal to the new foliation 
only play an auxiliary role.

The relevant change of foliation is made by switching to 
new coordinates $r{\equiv}a(x^0){\sinh}{\chi}$, 
${x^0}'{\equiv}a(x^0){\cosh}{\chi}$, so that the line element (2) transforms to
\eqa
{ds}^2&=&-{\Big (}{\frac{{\dot a}^{-2}-{\frac{r^2}{({x^0}')^2}}}{1-
{\frac{r^2}{({x^0}')^2}}}}{\Big )}(d{x^0}')^2
-2{\frac{r}{{x^0}'}}{\Big (}{\frac{1-{\dot a}^{-2}}{1-
{\frac{r^2}{({x^0}')^2}}}}{\Big )}d{x^0}'dr \nonumber \\
&&+{\Big (}{\frac{1-{\frac{r^2}{({x^0}')^2}}{\dot a}^{-2}}{1-
{\frac{r^2}{({x^0}')^2}}}}{\Big )}dr^2+r^2d{\Omega^2}, \qquad
{\frac{r}{{x^0}'}}<{\inf}
{\{}{\mid}{\dot a}{\mid}^{-1},{\mid}{\dot a}{\mid}{\}},
\ena
where ${\dot a}{\equiv}{\frac{da}{dx^0}}={\dot a}({x^0}',r)$ is now a function
of both the time coordinate and the radial coordinate. We can now use equation
(8) to find the spectral shift of light emitted by a FO located at the radial 
coordinate ${\chi}$ (i.e., with coordinate motion ${\frac{dr}{d{x^0}'}}=
{\frac{r}{{x^0}'}}$) as observed by a FO located at the origin ${\chi}=0$. 
Moreover, for small values of ${\chi}$, we will show that to lowest order in 
${\frac{r}{{x^0}'}}$, this spectral shift can be written as a sum of 
``kinematic'' and ``gravitational'' contributions. Note that, since
$a(x^0)={\sqrt{({x^0}')^2-r^2}}={x^0}'+O(2)$, the choice of foliation (i.e.,
the choice of time coordinate) leading to equation (8) is unique to first order
in the small quantity ${\frac{r}{{x^0}'}}$, but not higher. This means that 
any split-up of spectral shifts into ``kinematic'' and ``gravitational'' parts 
is limited to small distances.

To arrive at the desired spectral shift split-up, we first split up the 
4-velocity ${\bf u}_{\cal F}$ of the FOs into parts normal and tangential to 
the hypersurfaces ${x^0}'=$constant. This split-up reads
\eqa
{\bf  u}_{\cal F}={\gamma}(c{\bf {\tilde n}}+{\bf w}_{\cal F}), \qquad
{\gamma}{\equiv}(1-{\frac{w_{\cal F}^2}{c^2}})^{-1/2},
\ena
where ${\bf {\tilde n}}$ is the unit normal vector field of the hypersurfaces 
${x^0}'=$constant and ${\bf w}_{\cal F}$ is the 3-velocity difference (with 
squared norm $w^2_{\cal F}$) between a FO and a local observer moving normal 
to these hypersurfaces. Only the $r$-component of this equation is of interest,
and it reads
\eqa
w^r_{\cal F}={\frac{dr}{d{x^0}'}}{\frac{d{x^0}'}{d{\tau}_N}}+{\frac{N^r}{N}}c,
\quad cd{\tau}_N=Nd{x^0}'={\sqrt{
{\frac{1-{\frac{r^2}{({x^0}')^2}}}{{\dot a}^{2}-{\frac{r^2}{({x^0}')^2}}}}}
}d{x^0}', \quad
N^r={\frac{r}{{x^0}'}}{\frac{1-{\dot a}^{2}}{[{\dot a}^2-
{\frac{r^2}{({x^0}')^2}}]}},
\ena
where $N$ is the lapse function and $N^r$ is the shift vector $r$-component
of observers moving normal to the hypersurfaces ${x^0}'=$constant. The 
relation of these quantities to the line element given by equation (8) can be 
found from the formula
\eqa
{ds}^2=[N^rN_r-N^2](d{x^0}')^2+2N_rd{x^0}'dr+{\tilde h}_{ij}dx^idx^j,
\ena
where ${\tilde h}_{ij}$ are the components of the hypersurface metric, found 
explicitly from equation (8). It is straightforward to calculate the speed
$w_{\cal F}$, and we find that
\eqa
w_{\cal F}{\equiv}{\sqrt{w^i_{\cal F}w^j_{\cal F}{\tilde h}_{ij}}}=
{\frac{{\tanh}{\chi}}{{\mid}{\dot a}{\mid}}}c
={\frac{r}{{\mid}{\dot a}{\mid}{x^0}'}}c.
\ena
The speed $w_{\cal F}$ can now be put into the special-relativistic Doppler 
formula to find the spectral shift as measured by a local observer moving 
normal to the hypersurfaces ${x^0}'=$constant. Applied to the Milne model,
this approach yields a local determination of the cosmological spectral shift
in flat space-time. It is thus natural to define a more general ``kinematic'' 
spectral shift $z_k$ valid for open models, found locally and given by
\eqa
1+z_k{\equiv}{\sqrt{{\frac{1{\pm}w_{\cal F}/c}{1{\mp}w_{\cal F}/c}}}},
\qquad \Rightarrow \qquad
z_k={\pm}{\frac{r}{{\mid}{\dot a}{\mid}{x^0}'}}+O(2)
={\dot a}^{-2}{\frac{H(x^0)r}{c}}+O(2).
\ena
We see that if ${\mid}{\dot a}{\mid}{\rightarrow}1$, we have a situation where 
the spectral shift is defined as ``purely kinematic'' according to 
Definition 2, and is identical to the special-relativistic result. Moreover,
one may easily see that this definition yields the expected result $z_k=0$ if 
extrapolated to RW-models with flat spatial sections.  That is, the coordinate 
transformation $r{\equiv}a(x^0){\chi}$, ${x^0}'{\equiv}a(x^0)$ yields the
counterpart expression to equation (8) of the line element valid for RW-models
with flat space sections. Hence $w_{\cal F}{\equiv}0$ due to the fact that 
this coordinate transformation does not yield a new foliation so that the FOs 
move orthogonally to the ${x^0}'=$constant hypersurfaces as well.

Next, we note that an observer moving with constant $r$-coordinate and a local 
observer moving normal to the hypersurfaces ${x^0}'=$constant will not have 
coinciding world lines, but will have a 3-velocity difference ${\bf w}$. We 
will now show that the corresponding speed $w$ can be used to define a local 
determination of ``gravitational'' spectral shift. To do that, similar to 
equations (10) and (12), we find the quantities
\eqa
w^r={\frac{N^r}{N}}c={\frac{rc}{{x^0}'}}{\frac{1-{\dot a}^2}{
{\sqrt{({\dot a}^2-{\frac{r^2}{({x^0}')^2}})(1-{\frac{r^2}{({x^0}')^2}})}}}},
\quad \Rightarrow \quad
w={\frac{{\mid}1-{\dot a}^2{\mid}r}{{\mid}{\dot a}{\mid}{x^0}'(1-
{\frac{r^2}{({x^0}')^2}})}}c,
\ena
and these expressions vanish in the limit ${\mid}{\dot a}{\mid}{\rightarrow}1$,
as they should for ``gravitational'' quantities. It is thus natural to 
associate the corresponding spectral shift with space-time curvature,
i.e., it should be due to ``gravitational'' causes. Since the sign of $w^r$ 
depends on whether $1-{\dot a}^2$ is positive or negative, the contribution to 
the spectral shift with respect to the emitting FO will be either negative or 
positive, respectively. That is, what enters into the special-relativistic 
Doppler formula is not the speed $w$, but rather the quantity $w_{\pm}$ 
defined by
\eqa
w_{\pm}{\equiv}{\frac{({\dot a}^2-1)r}{{\mid}{\dot a}{\mid}{x^0}'(1-
{\frac{r^2}{({x^0}')^2}})}}c,
\ena 
which may be used to define a ``gravitational'' spectral shift $z_g$ 
(valid for open models) given by
\eqa
1+z_g{\equiv}{\sqrt{{\frac{1{\pm}w_{\pm}/c}{1{\mp}w_{\pm}/c}}}}, 
\qquad \Rightarrow \qquad
z_g={\pm}{\frac{r({\dot a}^2-1)}{{\mid}{\dot a}{\mid}{x^0}'}}+O(2).
\ena
Again, one may easily check that a similar definition extrapolated to RW-models
with flat spatial sections yields the expected lowest-order result 
$z_g={\frac{H({x^0}')r}{c}}$.

If the observer moving with constant $r$-coordinate emits light that is
detected by the FO residing in the spatial origin, the resulting spectral
shift will be of higher order in the small quantity ${\frac{r}{{x^0}'}}$, so
this contribution can be neglected. (Here, a possible effect of nonzero
${\ddot a}{\equiv}{\frac{d^2a}{{dx^0}^2}}$ may also be neglected if said small
quantity is small enough.) This means that to lowest order, the total spectral 
shift measured by the FO residing at the origin can be written as a sum of 
``kinematic'' and ``gravitational'' contributions, and that this spectral 
shift is given by
\eqa
z&=&z_k+z_g+O(2)={\pm}{\Big [}({\dot a}^2-1)+1{\Big ]}
{\frac{r}{{\mid}{\dot a}{\mid}{x^0}'}}+O(2) \nonumber \\
&&={\pm}{\mid}{\dot a}{\mid}
{\frac{r}{{x^0}'}}+O(2)={\frac{H(x^0)r}{c}}+O(2),
\ena
which is the familiar lowest-order expression for cosmological spectral shifts.
Moreover, for small distances the split-up defined in equation (17) is unique.
On the other hand, for large distances, cosmological spectral shifts in an
open RW-model cannot uniquely be split up into ``kinematic'' and 
``gravitational'' parts. This is so since other foliations (coinciding with
the foliation defined by the ${x^0}'$-coordinate for small distances but 
differing from it for large distances) may be equally well be used when 
defining spectral shift split-up by the method described above.

To illustrate the meaning of the split-up defined in equation (17), we finish 
this section with some simple examples. First, we choose a form of the scale 
factor consistent with a radiation-dominated universe as predicted by GR, i.e.,
\eqa
a(x^0)={\sqrt{a^*x^0}}, \quad 
{\dot a}={\frac{1}{2}}{\sqrt{\frac{a^*}{x^0}}}={\frac{a^*}{2a}}, \quad 
z_k={\frac{2r}{a^*}}+O(2), \quad z_g=[{\frac{1}{2x^0}}-{\frac{2}{a^*}}]r+O(2),
\ena
where $a^*$ represents an arbitrary constant reference scale. We note that
$z_k$ does not depend on epoch. Furthermore, we see that $z_g$ is positive for 
early epochs, vanishes for $x^0=a^*/4$, and becomes negative for later epochs. 
The particular epoch where $z_g$ vanishes is (of course) determined by the 
condition ${\dot a}=1$. At this epoch, the expansion of the universe 
momentarily mimics that of the Milne model with a ``shifted'' scale factor 
given by $a(x^0)=x^0+a^*/4$. Hence, since we can neglect the effect of 
${\ddot a}$ for small enough distances, in this limit the ``kinematic'' 
interpretation of the cosmic redshift will hold, despite the fact that 
space-time is not flat. However, at earlier epochs ${\dot a}>1$ so the universe 
expands faster than an ``empty'' universe, yielding an extra redshift. This
means that gravity has not had enough time to slow down the expansion 
sufficiently over the universe's history. Similarly, for later epochs, 
${\dot a}<1$ and the universe expands slower than an ``empty'' universe, giving 
an extra blueshift. Gravity has then had enough time to slow down the 
expansion sufficiently so that it expands slower than in the Milne model.

Second, choose a form of the scale factor consistent with a matter-dominated
universe as predicted by GR, i.e.,
\eqa
a(x^0)=(a^*)^{\frac{1}{3}}(x^0)^{\frac{2}{3}}, \qquad \Rightarrow \qquad 
{\dot a}={\frac{2}{3}}{\Big (}{\frac{a^*}{x^0}}{\Big )}^{\frac{1}{3}}=
{\frac{2}{3}}{\sqrt{\frac{a^*}{a}}}, \nonumber \\
z_k={\frac{3r}{2(a^*)^{\frac{2}{3}}(x^0)^{\frac{1}{3}}}}+O(2), \quad 
z_g={\Big [}{\frac{2}{3x^0}}-
{\frac{3}{2(a^*)^{\frac{2}{3}}(x^0)^{\frac{1}{3}}}}{\Big ]}r+O(2),
\ena
where again $a^*$ is an arbitrary constant reference scale. We note that, 
unlike the previous example, in this case $z_k$ depends on epoch. Moreover, 
$z_g$ vanishes for the epoch $x^0={\frac{8}{27}}a^*$, and at this epoch the 
universe expands momentarily as an ``empty'' RW-model with a ``shifted'' scale 
factor given by $a(x^0)=x^0+{\frac{4}{27}}a^*$. So at this particular epoch
the cosmic redshift should be interpreted as a pure ``kinematic'' effect in 
flat space-time for small distances (even though space-time is not flat). 
However, this interpretation breaks down for other epochs.

A final example is given where the scale factor is determined by a (positive)
cosmological constant ${\Lambda}$, i.e.,
\eqa
a(x^0)={\sqrt{\frac{3}{\Lambda}}}{\sinh}{\Big [}
{\sqrt{\frac{{\Lambda}}{3}}}x^0{\Big ]}, 
\qquad \Rightarrow \qquad 
{\dot a}={\cosh}{\Big [}{\sqrt{{\frac{{\Lambda}}{3}}}}x^0{\Big ]}=
{\sqrt{1+{\frac{{\Lambda}}{3}}a^2}}, \nonumber \\
z_k={\frac{r}{a{\sqrt{1+{\frac{\Lambda}{3}}a^2}}}}+O(2), \qquad 
z_g={\frac{\Lambda}{3}}{\frac{ar}{{\sqrt{1+{\frac{\Lambda}{3}}a^2}}}}+O(2).
\ena
We note that in this case, for very early epochs $x^0{\rightarrow}0$, the 
cosmic expansion mimics that of the Milne model so that 
$z_k{\rightarrow}{\infty}$ and $z_g{\rightarrow}0$ in this limit. However, at 
late epochs $z_k$ decreases exponentially, so it can soon be neglected. Thus, 
at late epochs, the cosmic redshift should be interpreted as due to space-time 
curvature (i.e., ``gravitational'') with negligible ``kinematic'' contribution.
\section{Fallacies of popular cosmology}
The results obtained in section 2 for the flat and closed RW-models were also 
arrived at by Roukema [1], using topological methods. That is, by changing the
topology of the spatial sections of the relevant metrics (given by equations 
(2) and (3)) from simply connected to multiply connected, but without changing 
the geometry, it was shown that a contradiction arises if spectral shifts are 
interpreted as due to the Doppler effect in flat space-time. On the other hand,
considering a less general case than for flat and closed RW-models, this 
contradiction did not occur for open RW-models, except for certain large 
distances. This means that the results obtained in section 3 do not match the
corresponding results in [1], so there seems to be a contradiction. (This would
indicate that using topological methods is not sufficient for analysing the
RW-models with hyperbolic spatial sections.) On the other hand, searching the 
relevant literature, one finds that reference [1] is about the only one 
emphasizing the crucial role of the spatial geometry when it comes to 
interpretations. Otherwise, what has been discussed is the ``kinematic'' 
versus the ``expanding space'' views with no due weight on spatial geometry. 
It has even been claimed [5] that spatial geometry is irrelevant for 
interpretations of certain cosmological {\em gedanken}-experiments involving 
radar distances and spectral shifts, since the calculated results of such 
hypothetical experiments do not depend on the spatial parts of the metrics (2).
But this argument is flawed since the actual debate is about interpretations of 
{\em models} rather than of experimental results. Moreover, there is absolutely 
no scientific requirement that different interpretations of models should be 
experimentally distinguishable.

Another, common but incorrect assertion is that the effects on spectral shifts
of curved space-time, as compared to ``kinematic'' effects,  can always
be neglected in the RW-models for sufficiently small distances. The argument is
that, since one may always choose local coordinates such that the tangent 
space-time at some event ${\cal P}$ (given, e.g., by $x^0=x^0_0$, ${\chi}=0$) 
takes the standard Minkowski form, and in a (small) neighbourhood of 
${\cal P}$ approximates the space-time metric to first order in small 
quantities, the effects of space-time curvature can be made negligible in a 
sufficiently small neighbourhood of ${\cal P}$. (The EP ensures that such a 
coordinate system can be found for {\em any} metric.) So far, the argument is 
of course correct. But it is then incorrectly claimed that in such a coordinate
system, the FOs will define a (radial) velocity field $v(r)=H_0r+O(2)$ (where 
$H_0$ is the local Hubble parameter) with respect to the FO momentarily 
residing in ${\cal P}$ (where $r=0$). Since to desired accuracy, $v(r)$ by 
construction represents space-time geodesics defining a velocity field in flat 
space-time, it is concluded that this proves that cosmic spectral shifts
must be interpreted as a purely ``kinematic'' for small enough distances. 

The flaw in this reasoning is that the inertial observers defining $v(r)$
can in general not be identified with the FOs. That is, it is certainly always 
possible to construct a set of geodesics in flat space-time defining a velocity
field ${\bf v}$ with respect to some chosen specific observer, such that 
photon signalling between this chosen observer and the observers defining 
${\bf v}$ mimics Hubble spectral shifts. It is also always possible to identify 
the chosen observer with some FO in a curved RW-manifold. But there is 
absolutely no guarantee that the FOs in the curved RW-manifold can be 
identified with the observers defining ${\bf v}$. Such observers will in 
general be {\em some other} observers, moving along {\em different} geodesics 
than the FOs. In other words, in the curved RW-manifold one started out with, 
these other observers will have non-zero peculiar velocities with respect to 
the FOs. This will obviously not change in the flat space-time approximation, 
since geodesic deviation can be neglected for small enough regions. This means
that, to be able to interpret the velocity field ${\bf v}$, it is necessary
to know the relationship between the FOs and the observers defining ${\bf v}$.
That relationship can only be found from the nature of the connection, i.e.,
by knowing how well it may be approximated by a flat connection. For example,
for the situation described in Definition 2, a flat connection is a sufficient
approximation so that the FOs really can be identified with the observers 
defining ${\bf v}$. On the other hand, a flat connection contributes nothing 
at all to ${\bf v}$ for RW-models with flat or spherical space sections.

To see that the effects of a non-flat connection cannot be neglected in general,
even for small distances, it is illustrating to write the scale factor 
$a(x^0)$ as a Taylor series around the event ${\cal P}$, i.e., 
\eqa
a(x^0)=a(x^0_0)+{\dot a}(x^0_0)[x_0-x^0_0]
+{\frac{1}{2}}{\ddot a}(x^0_0)[x^0-x^0_0]^2+{\cdots}, \qquad 
v(r)={\frac{{\dot a}(x^0_0)}{a(x^0_0)}}rc+O(2).
\ena
Since the relevant connection coefficient for radial motion as obtained from
equation (2) is given by ${\Gamma}^{\chi}_{0{\chi}}={\frac{{\dot a}}{a}}$, we 
see that the construction of the velocity field $v(r)$ in flat space-time 
depends only on the fact that this connection coefficient is non-zero. Since 
this is true regardless of the RW-model, one has actually by construction 
transformed {\em all} relevant effects, ``kinematic'' and curvature effects 
alike, into $v(r)$. In other words, since nothing at all is said regarding the 
nature of ${\Gamma}^{\chi}_{0{\chi}}$, and the observers defining $v(r)$ remain 
unidentified, the construction of $v(r)$ is in fact {\em irrelevant} for 
interpretations of the expansion.

A paper based on the faulty line of reasoning outlined above is [6], claiming 
that interpretations of spectral shifts between FOs for small distances 
depend on the choice of coordinate system and method of calculation. Moreover,
it is argued that cosmological spectral shifts are most ``naturally'' 
interpreted as Doppler shifts in flat space-time for small distances. But as 
we have seen, these claims are simply incorrect. A related idea advocated in 
[6], is that spectral shifts between FOs can equally ``naturally'' be 
interpreted as Doppler shifts in flat space-time even for large distances. To 
justify this assertion, the total spectral shift is being thought of as an 
accumulated effect of many small Doppler effects in flat space-time. But this 
logic will, of course, break down since spectral shifts between FOs cannot, in 
general, consistently be interpreted as Doppler shifts in flat space-time even 
for small distances. On the other hand, the antithesis of [6] is a paper [7] 
where it is (also incorrectly) argued that cosmic spectral shifts involving 
FOs only must ``definitely'' be interpreted as ``gravitational'' (with an 
exception for the Milne model). This claim is based on the specific choice of
(discontinuous) scale factor $a(x^0)=1+{\theta}(x^0)$, where ${\theta}(x^0)$ is
the Heaviside step function. It is then argued that the resulting cosmological
spectral shift cannot be interpreted  as a Doppler shift in flat space-time 
since both source and receiver are at rest when the signal is emitted or 
received. Moreover, it is argued that the sudden ``non-local motion'' occurring
in this example should shed light on the interpretation of cosmological 
spectral shifts obtained in any ``non-empty'' RW-model. However, as we have 
seen in sections 2 and 3, a mere choice of scale factor without considering 
spatial geometry is not sufficient for interpretations of cosmological spectral 
shifts obtained in the RW-models. Besides, counterexamples to the claim that
cosmological spectral shifts obtained from any non-empty RW-model must 
``definitely'' be interpreted as ``gravitational'' are presented in section 3 
of this paper, for situations more generally described in Definition 2 (see 
section 2).

There have also been earlier attempts to split up cosmic spectral shifts into 
``kinematic'' and ``gravitational'' parts (for small distances). Such have 
been based on a Taylor expansion similar to that shown in equation (21) in 
combination with a Newtonian approximation to calculate the ``gravitational'' 
contribution (a second order blueshift, see, e.g., [8, 9]). However, as we 
have seen, using equation (21) for this purpose is misguided. Besides, since 
interpretations of spectral shifts in the RW-models should be based on their
geometric properties only, without referring to specific dynamical laws, 
any use of Newtonian approximations only confuses the issue.  

A recent attempt of defining said split-up in general (even for large 
distances) has been made in [10]. In that paper, the  ``recession velocity'' is 
defined as the 3-velocity obtained by parallel-transporting the 4-velocity of 
the emitting FO to the observing FO along a space-like geodesic lying in a 
hypersurface of constant cosmic time, and then projecting the resulting 
4-velocity into the local rest frame of the observing FO. This ``recession 
velocity'' then defines the ``kinematic'' part of the cosmic spectral shift. 
(But as shown in section 2, this approach does not make sense for RW-models 
with flat or spherical space sections since with this definition, there is no 
correspondence with spectral shifts in SR.) The difference between the total 
cosmic spectral shift and the ``kinematic'' spectral shift is interpreted as a 
``gravitational'' spectral shift. It was shown that this definition of 
``gravitational'' spectral shift agrees with that found in [8, 9] for small 
distances. But while the effort made in [10] is certainly ingenious, this does 
not change the fact that the resulting interpretations are in general 
inconsistent with the geometry of the RW-models, as explained in this paper 
and in [1].
\section{Conclusion}
For several years, a debate has been going on in the scientific literature 
regarding the nature and interpretation of cosmological spectral shifts. This 
debate is primarily about theoretical models based on GR and whether or not 
different interpretations of cosmic spectral shifts are consistent with these
models. 

As a general rule, any interpretation is consistent with any theoretical model 
as long as no logical or mathematical inconsistencies arise. Therefore, 
different interpretations of the same features of any model may in principle 
be possible. One often hears that this is the case for the interpretation of 
cosmic spectral shifts. However, in this paper, we have shown that the school 
claiming general validity (at least for sufficiently small distances) of a 
``kinematic'' interpretation of cosmological spectral shifts, is in error. This
is so since geometric properties of the RW-models are inconsistent with 
such interpretations, except for the Milne model and special epochs in open 
RW-models. In particular, we have shown that for flat and closed RW-models,
there can be no cosmic expansion without the relevant space-time curvature 
since otherwise, the Gauss equation would be violated. Therefore, in these 
models, cosmic spectral shifts must be interpreted as an effect solely due to 
space-time curvature. For open models, interpretations are more subtle, since
here, at least part of the spectral shifts will be ``kinematic''.

So, is the nature of the cosmic expansion now fully understood and all
controversy settled once and for all? This is not likely, since convincing
opponents of their erroneous arguments and points of view is very difficult. 
Besides, an alternative space-time framework exists, the so-called quasi-metric
framework (QMF), where the cosmic expansion is described as new physics not
covered by GR or Newtonian concepts, and its nature differs radically from its 
counterpart in the RW-models [3]. The QMF describes the nature of the cosmic 
expansion as ``non-kinematic'' in the sense that it is not a part of 
space-time's causal structure. (Thus one may argue that the nature of the 
cosmic expansion in the RW-models is indeed ``kinematic'' in the broader sense 
of being part of space-time's causal structure, without specifying any 
particular dynamical model.) Moreover, unlike GR, quasi-metric space-time
is by construction equipped by a ``preferred'' global foliation into 
3-dimensional, simply connected and closed spatial hypersurfaces defining 
``space''. There is some resemblance to a closed RW-model since the 
``non-kinematic'' expansion also defines extra space-time curvature via a 
non-flat connection. That is, just as for the closed RW-models, in the QMF the 
cosmic redshift is an effect of space-time curvature. However, a rather unique 
prediction of the QMF is that gravitationally bound systems should expand in 
general, and this prediction has observational support in the solar system 
[11, 12]. (However, the significance of the observations referred to in [12]
has been challenged in recent years.) This means that it should be possible in 
principle to test the nature of the cosmic expansion by doing controlled 
experiments in the solar system. But based on the GR prediction that the cosmic
expansion in the solar system should be far too small to be detectable, both 
the evidence in favour of local cosmic expansion and the possibility of doing 
controlled experiments to test it have been ignored so far. 

As a final remark, I regret to say that if the scientific discussion regarding
popular cosmology were sound, it would not have been necessary for me to write
the present paper. However, in this field much low-quality and confusing
material has been published by people who should know better. As a result, 
several incorrect arguments based on personal intuition seem to have been 
accepted as ``mainstream'', misleading people and in particular students. Such 
breach of decent scholarship cannot be allowed to pass without notice.
\\ [4mm]
{\bf References} \\ [1mm]
{\bf [1]} B.F. Roukema, {\em MNRAS} {\bf 404}, 318 (2010) (arXiv:0911.1205). \\
{\bf [2]} J.V. Narlikar, {\em Am. Journ. Phys.} {\bf 62}, 903 (1994). \\
{\bf [3]} D. {\O}stvang, {\em Gravitation {\&} Cosmology} {\bf 11}, 205 (2005)
(gr-qc/0112025). \\
{\bf [4]} R.M. Wald, {\em General Relativity}, The University of 
Chicago Press (1984). \\
{\bf [5]} M.A. Abramowicz, S. Bajtlik, J-P. Lasota and A. Moudens, \\
{\hspace*{6.4mm}}{\em Acta Astron.} {\bf 57}, 139 (2007) (astro-ph/0612155); \\
{\hspace*{6.4mm}}{\em Acta Astron.} {\bf 59}, 131 (2009) (arXiv:0812.3266). \\
{\bf [6]} E.F. Bunn and D.W. Hogg, {\em Am. Journ. Phys.} {\bf 77}, 688 (2009)
(arXiv:0808.1081). \\
{\bf [7]} V. Faraoni, {\em Gen. Rel. Grav.} {\bf 42}, 851 (2010) 
(arXiv:0908.3431). \\
{\bf [8]} H. Bondi, {\em MNRAS} {\bf 107}, 410 (1947). \\
{\bf [9]} {\O}. Gr{\o}n and {\O}. Elgar{\o}y, {\em Am. Journ. Phys.} {\bf 75},
151 (2007) (astro-ph/0603162). \\
{\bf [10]} M.J. Chodorowski, {\em MNRAS} {\bf 413}, 585 (2011)
(arXiv:0911.3536). \\
{\bf [11]} D. {\O}stvang, {\em Gravitation {\&} Cosmology} {\bf 13}, 1 (2007)
(gr-qc/0201097). \\
{\bf [12]} D. {\O}stvang, {\em Class. Quantum Grav.} {\bf 19}, 4131 (2002)
(gr-qc/9910054). 
\end{document}